\documentstyle[proceedings,numreferences]{crckapb}
 
\begin{opening}
 
\title{Renormalization constants \protect \\ from string theory} 
 
\author{Paolo Di Vecchia}
\author{Lorenzo Magnea}
\institute{NORDITA \\ Blegdamsvej 17, DK-2100 Copenhagen \O, Denmark}
  
\author{Alberto Lerda}
\author{Rodolfo Russo}
\institute{Dipartimento di Fisica Teorica, Universit\`a di Torino \\
Via P.Giuria 1, I-10125 Turin, Italy \\
and I.N.F.N., Sezione di Torino}
 
\author{Raffaele Marotta}
\institute{Dipartimento di Scienze Fisiche, Universit\`a di Napoli \\
Mostra D'Oltremare, Pad. 19, I-80125 Napoli, Italy}

\end{opening} 

\begin{document}


\newcommand{\secn}[1]{Section~\ref{#1}}
\newcommand{\bra}[1]{\langle{#1}|}
\newcommand{\ket}[1]{|{#1}\rangle}
\newcommand{\braket}[2]{\langle{#1}|{#2}\rangle}
\newcommand{\tbl}[1]{Table~\ref{#1}}
\newcommand{\eq}[1]{Eq.~(\ref{#1})}
\newcommand{\fig}[1]{Fig.~\ref{#1}}
\newcommand{\nl}{\nonumber \\}
\def\zeroslash{\mathord{\not\mathrel{\hskip 0.04 cm 0}}}
\def\beq{\begin{equation}}
\def\eeq{\end{equation}}
\def\beqa{\begin{eqnarray}}
\def\eeqa{\end{eqnarray}}

\newcommand{\sect}[1]{\setcounter{equation}{0}\section{#1}}
\renewcommand{\theequation}{\thesection.\arabic{equation}}
\newcommand{\EQ}{\begin{equation}}
\newcommand{\EN}{\end{equation}}
\newcommand{\bea}{\begin{eqnarray}}
\newcommand{\ena}{\end{eqnarray}}
\newcommand{\vs}[1]{\vspace{#1 mm}}
\renewcommand{\a}{\alpha}
\renewcommand{\b}{\beta}
\renewcommand{\c}{\gamma}
\renewcommand{\d}{\delta}
\newcommand{\e}{\epsilon}
\newcommand{\ve}{\varepsilon}
\newcommand{\hs}[1]{\hspace{#1 mm}}
\newcommand{\shalf}{\frac{1}{2}}
\renewcommand{\Im}{{\rm Im}\,}
\newcommand{\NP}[1]{Nucl.\ Phys.\ {\bf #1}}
\newcommand{\PL}[1]{Phys.\ Lett.\ {\bf #1}}
\newcommand{\NC}[1]{Nuovo Cimento {\bf #1}}
\newcommand{\CMP}[1]{Comm.\ Math.\ Phys.\ {\bf #1}}
\newcommand{\PR}[1]{Phys.\ Rev.\ {\bf #1}}
\newcommand{\PRL}[1]{Phys.\ Rev.\ Lett.\ {\bf #1}}
\newcommand{\MPL}[1]{Mod.\ Phys.\ Lett.\ {\bf #1}}
\newcommand{\IJMP}[1]{Int.\ Jour.\ of\ Mod.\ Phys.\ {\bf #1}}
\renewcommand{\thefootnote}{\fnsymbol{footnote}}



\renewcommand{\baselinestretch}{1.0}
\textwidth 149mm
\textheight 220mm
\topmargin -.5in
\oddsidemargin 5mm

 
\begin{abstract}

We review some recent results on the calculation of renormalization
constants in Yang-Mills theory using open bosonic strings.
The technology of string amplitudes, supplemented with an appropriate
continuation off the mass shell, can be used to compute the ultraviolet
divergences of dimensionally regularized gauge theories. The results
show that the infinite tension limit of string amplitudes corresponds
to the background field method in field theory.
(Proceedings of the Workshop ``Strings, Gravity and Physics at the
Planck scale '', Erice (Italy), September 1995. Preprint DFTT 82/95)
\end{abstract}
  
\sect{Introduction}
\label{intro}

In the past few years it has become clear that string theory is
not only a good candidate for a unified theory of all interactions, but
also a useful tool to understand the structure of perturbative field
theories. Field-theoretical results can be recovered from string theory
by decoupling the infinite tower of massive string states, that is by
taking the limit of infinite string tension, or equivalently of vanishing
Regge slope $\a'$.

Since in the limit $\alpha ' \rightarrow 0$ string theories reduce to 
non-abelian gauge theories, unified with gravity, order by order in 
perturbation theory, in this limit we may expect to reproduce, order 
by order, scattering amplitudes, ultraviolet divergences,
and other physical quantities that one computes perturbatively in
non-abelian gauge theories.

A very useful feature of string theory for this purpose is the fact that, 
at each order of string perturbation theory, one does not get 
the large number of diagrams characteristic of field theories,
which makes it very difficult to perform high order calculations. 
Using closed strings, one gets only one diagram at each order, 
while with open strings the number of diagrams remains limited. 
Furthermore, compact expressions for these diagrams are known 
explicitly for an arbitrary perturbative order~\cite{copgroup}, in contrast
with the situation in field theory, where no such all-loop formula is 
known. Finally, string amplitudes are naturally written 
in a way that takes maximal advantage of gauge invariance: the color 
decomposition is automatically performed, and so are integrations over 
loop momenta, so that the helicity formalism is readily implemented.

The combination of these different features of string theory has led
several authors~\cite{mets,mina,tayven,kaplu,berkosbet,berkoswfr} to use
string theory as an efficient conceptual and computational 
tool in different areas of perturbative field theory. 
In particular, because of the compactness of the multiloop string expression, 
it is in some cases easier to calculate non-abelian gauge theory amplitudes
by starting from a string theory, and performing the zero slope 
limit, rather than using traditional techniques. 
In this way the one-loop amplitude involving four external gluons has been 
computed, reproducing the known field-theoretical result with much less 
computational cost~\cite{berkos}. Following the same approach, also 
the one-loop five gluon amplitude has been computed for the first 
time~\cite{fiveglu}.

The aim of this talk is to summarize the results obtained in 
Refs.~\cite{letter} and~\cite{paper}. There it was shown that, provided
a simple off-shell continuation is performed, string theory also contains
information on the ultraviolet divergences of Yang-Mills theory, and the
information can be consistently extracted in the language of dimensional 
regularization. In particular, starting from the one-loop two, three and 
four-gluon amplitudes in the open bosonic string, we performed the field 
theory limit, and we showed that in this limit
the renormalization constants $Z_A ,Z_3 $ and $Z_4$ of non-abelian gauge 
theories can be consistently recovered. String theory leads unambiguously 
to the background field method, as suggested by the on-shell analysis
of Ref.~\cite{berdun}.

Before going into the details of the calculation, let us first recall how 
field theory amplitudes are obtained from string theory, and how we 
expect those amplitudes to be renormalized.

In field theory one normally computes either connected Green functions,
denoted here by $W_{M} (p_1 \dots p_M )$, or one-particle irreducible (1PI) 
Green functions, $\Gamma_M (p_1 \dots p_M )$. In both cases, in general, an
off-shell continuation is performed, in order to avoid possible infrared 
divergences.

In string theory, on the other hand, one computes $S$-matrix elements 
involving gluon states, which are connected, via the reduction formulas, 
to on-shell connected Green functions, truncated with free propagators. 
Taking the field theory limit, the natural ultraviolet regulator of 
string theory, $1/\a'$, is removed, so that the usual divergences are 
recovered. The Green functions one computes are thus unrenormalized, 
and a new regulator must be introduced, in our case dimensional continuation. 
We will see that also in this case an off-shell extrapolation is necessary 
in order to avoid infrared problems.
 
Once the field theory limit is taken, it is possible to isolate 1PI 
contributions, which lead to the 1PI Green functions $\Gamma_M$, or to
compute the full amplitudes, which lead to the Green functions $W_M$.
{}From the knowledge on how they renormalize we can then extract the
renormalization constants. For example,
\beq
\Gamma_2 (g) = Z_{A}^{-1} \Gamma^{(R)}_{2} (g) \hspace{1cm}
\Gamma_3 (g) = Z_{3}^{-1} \Gamma^{(R)}_{3} (g) \hspace{1cm}
\Gamma_4 (g) = Z_{4}^{-1} \Gamma^{(R)}_{4} (g)~~~, 
\label{renprop1}
\eeq
while
\beq
W_3 (g) = Z_{3}^{-1} Z_{A}^{3} W_{3}^{(R)} (g)~~~,
\label{renprop2}
\eeq
where $g$ is the renormalized coupling constant.

The talk is organized as follows. In \secn{mglue} we consider the open 
bosonic string, and we write the explicit expression of the $M$ gluon 
amplitude at $h$ loops, including the overall normalization. In \secn{tree} we 
give the relevant amplitudes for the  tree and one-loop diagrams. 
In \secn{twopoint} we sketch  the calculation of the one-loop two gluon 
amplitude, already presented in~\cite{letter}, and we extract 
the gluon wave function renormalization constant $Z_A$. In \secn{effact} we 
present an alternative method, that allows one to exactly integrate
over the punctures, and we use it to extract the renormalization constants 
$Z_A$, $Z_3$ and $Z_4$. Finally, in \secn{threepoint} we extend the 
calculation of \secn{twopoint} to the one-loop three gluon amplitude, and 
we discuss how to extract the contribution of the one-particle reducible 
diagrams, that were neglected in \secn{effact}. \secn{concl} contains 
concluding remarks.

\sect{The $M$-gluon $h$-loop amplitude}
\label{mglue}
 
In string theory the $M$-gluon scattering amplitude can be computed 
perturbatively and is given by
\bea
A(p_1, \dots , p_M)&=&\sum_{h=0}^\infty
A^{(h)}(p_1, \dots , p_M) \nl
&=& \sum_{h=0}^\infty g_s^{2h-2}
{\hat A}^{(h)}(p_1, \dots , p_M)~~~,
\label{pertampl}
\ena
where $g_s$ is a dimensionless string coupling constant, which is
introduced to formally control the perturbative expansion.
In \eq{pertampl}, $A^{(h)}$ represents the $h$-loop contribution.
For the closed string $A^{(h)}$ is given by only one diagram,
while for the open string the number of diagrams is small in comparison
with the large number of diagrams encountered in field theory.

Let us consider the open bosonic string, and let us restrict ourselves
only to  planar diagrams. For such diagrams the $M$-gluon $h$-loop
amplitude, including the appropriate Chan-Paton factor, is given by
\bea
A^{(h)}_P (p_1,\ldots,p_M) & = & N^h\,{\rm Tr}(\lambda^{a_1}
\cdots \lambda^{a_M})\,
C_h\,{\cal N}_0^M    \nl  
& \times & \int [dm]^M_h \left\{\prod_{i<j} 
\left[{{\exp\left({\cal G}^{(h)}(z_i,z_j)\right)}
\over{\sqrt{V'_i(0)\,V'_j(0)}}}\right]^{2\a' p_i\cdot p_j} \right.  \nl
& \times & \exp\left[\sum_{i\not=j}
\sqrt{2\a'} p_j\cdot\ve_i 
\,\partial_{z_i} {\cal G}^{(h)}(z_i,z_j) \right.  \label{hmaster}  \\
& & + \, {1\over 2}\sum_{i\not=j} 
\ve_i\cdot\ve_j \,\partial_{z_i}\partial_{z_j}
{\cal G}^{(h)}(z_i,z_j)\Big]\Bigg\}_{\rm m.l.}~~~,  \nonumber
\ena
where the subscript ``m.l.'' stands for multilinear, meaning
that only terms linear in each polarization should be kept. 
\eq{hmaster} is written for transverse gluons, satisfying the condition
$\ve_i \cdot p_i = 0$, whereas the mass-shell condition $p_i^2 = 0$, though
necessary for conformal invariance of the amplitude, has not been enforced
yet.

The main ingredient in \eq{hmaster} is the $h$-loop world-sheet bosonic 
Green function ${\cal G}^{(h)}(z_i,z_j)$, which plays a key role in the
field theory limit. $[dm]^M_h$ is the measure of integration on moduli 
space for an open Riemann surface of genus $h$ with $M$ operator insertions
on the boundary~\cite{copgroup}.
The Green function ${\cal G}^{(h)}(z_i,z_j)$ can be expressed as
\beq
{\cal G}^{(h)}(z_i,z_j) = \log E^{(h)}(z_i,z_j) - {1\over 2} \int_{z_i}^{z_j} 
\omega^\mu \, \left(2\pi {\rm Im}\tau_{\mu\nu}\right)^{-1} 
\int_{z_i}^{z_j} \omega^\nu~~~, 
\label{hgreen}
\eeq
where $E^{(h)}(z_i,z_j)$ is the prime-form, $\omega^\mu$ ($\mu=1,\ldots, h$)
the abelian differentials and $\tau_{\mu\nu}$ the period
matrix of an open Riemann surface of genus $h$.
All these objects, as well as the measure on moduli space $[dm]^M_h$, can 
be explicitly written down in the Schottky parametrization of the Riemann 
surface, and their expressions for arbitrary $h$ can be found for example 
in Ref.~\cite{scho}. Here we will only write the explicit
expression for the measure, to give a flavor of the ingredients that enter 
the full string theoretic calculations. It is
\beqa
[dm]^M_h & = & \frac{\prod_{i=1}^M dz_i}{dV_{abc}}
\prod_{\mu=1}^{h} \left[ \frac{dk_\mu d \xi_\mu d \eta_\mu}{k_\mu^2
(\xi_\mu - \eta_\mu)^2} ( 1- k_\mu )^2 \right]   \label{hmeasure} \\
& \times & \left[\det \left( - i \tau_{\mu \nu} \right) \right]^{-d/2} 
\prod_{\alpha}\;' \left[ \prod_{n=1}^{\infty} ( 1 - k_{\alpha}^{n})^{-d}
\prod_{n=2}^{\infty} ( 1 - k_{\alpha}^{n})^{2} \right]~~~.   \nonumber
\eeqa
Here $\tau_{\mu \nu}$ is the period matrix, while $k_{\mu}$ are the 
multipliers and $\xi_{\mu}$ and $ \eta_{\mu}$ the fixed 
points of the generators of the Schottky group; $dV_{abc}$ is the projective 
invariant volume element 
\beq
dV_{abc} = \frac{d\rho_a~d\rho_b~d\rho_c}
{(\rho_a-\rho_b)~(\rho_a-\rho_c)~(\rho_b-\rho_c)}~~~,
\label{projvol}
\eeq
where $\rho_a$, $\rho_b$, $\rho_c$ are any three of the $M$
Koba-Nielsen variables, or of the $2h$ fixed points of the generators of the 
Schottky group, which can be fixed at will; finally, the primed product
over $\alpha$ denotes a product over classes of elements of the 
Schottky group~\cite{scho}.

Notice that in the open string the Koba-Nielsen variables must be cyclically 
ordered, for example according to
\beq 
z_1 \geq z_2 \cdots \geq z_{M}~~~,
\label{cyclord}
\eeq
and the ordering of Koba-Nielsen variables automatically
prescribes the ordering of color indices. 

The amplitude in Eq. (\ref{hmaster}) contains two normalization constants
which were calculated in Ref.~\cite{paper}, and are given by   
\EQ
C_h = {1\over{(2\pi)^{dh}}}~g_s^{2h-2}{1\over{(2\a')^{d/2}}} \hspace{1cm}
{\cal N}_0 = g_d \,\sqrt{2\a'}~~~,
\label{vertnorm}
\EN
where the string coupling $g_s$ and the $d$-dimensional gauge coupling $g_d$
are related by
\beq
g_s = \frac{g_d}{2}\,(2\a')^{1-d/4}~~~.
\label{gstring}
\eeq

An efficient way to explicitly obtain $A^{(h)}(p_1, \dots , p_M)$ 
is to use the $M$-point $h$-loop vertex $V_{M;h}$ of the operator
formalism. The explicit expression of $V_{M;h}$ for the 
planar diagrams of the open bosonic string can be found in 
Ref.~\cite{copgroup}. 
The vertex $V_{M;h}$ depends on $M$ real Koba-Nielsen variables $z_i$ 
through $M$ projective transformations $V_i(z)$, which define local 
coordinate systems vanishing around each $z_i$, {\it i.e.} such that 
\EQ
V_i^{-1}(z_i) = 0~~~.
\label{Vi}
\EN
When $V_{M;h}$ is saturated with $M$ physical string states satisfying 
the mass-shell condition, the corresponding amplitude does not depend on 
the $V_i$'s. However, as we discussed in Ref.~\cite{letter}, to extract
informations about the ultraviolet divergences that arise when the field 
theory limit is taken, it is necessary to relax the mass-shell condition, so 
that also the amplitudes $A^{(h)}$ will depend on the choice of projective 
transformations $V_i$'s, just like the vertex $V_{M;h}$. This is the reason of
the appearence of $V_i$ in Eq. (\ref{hmaster}).

\sect{Tree and one-loop diagrams}
\label{tree}

For tree-level amplitudes, corresponding to $h=0$, the situation is 
particularly simple. The Green
function in \eq{hgreen} reduces to 
\beq
{\cal G}^{(0)}(z_i,z_j) = \log (z_i-z_j)~~~,
\label{treegreen}
\eeq
while the measure $[dm]^M_0$ is simply
\beq
[dm]^M_0 = \frac{\prod\limits_{i=1}^M dz_i}{dV_{abc}}~~~.
\label{treemeas}
\eeq
Inserting Eqs. (\ref{treegreen}) and (\ref{treemeas}) into \eq{hmaster}, 
and writing explicitly all the normalization coefficients, we obtain
the color ordered, planar, on-shell $M$ gluon amplitude at tree level 
\bea
A^{(0)}_P (p_1,\ldots,p_M) & = & 4\,{\rm Tr}(\lambda^{a_1}
\cdots \lambda^{a_M})\,g_d^{M-2}\,(2\a')^{M/2-2}   \nl
& \times & \! \! \int_{\Gamma_0}\frac{\prod\limits_{i=1}^M dz_i}{dV_{abc}} 
\left\{\prod_{i<j} \left(z_i-z_j\right)^{2\a' p_i\cdot p_j} 
\right. \label{treemaster} \\
& \times & \left. \! \! \exp\left[\sum_{i<j}
\left(\sqrt{2\a'} \frac{p_j\cdot\ve_i 
-p_i\cdot\ve_j}{(z_i-z_j)} + \frac{\ve_i\cdot
\ve_j}{(z_i-z_j)^2}\right)\right]\right\}_{\rm m.l.}~, \nonumber
\ena
where $\Gamma_0$ is the region identified by \eq{cyclord}.
Notice that any dependence on the local coordinates $V_i(z)$
drops out in the amplitude after enforcing the mass-shell condition.
Notice also that \eq{treemaster} is valid only for $M \geq 3$, since the
measure given by \eq{treemeas} is ill-defined for $M \leq 2$.

We readily derive the three-gluon amplitude
\bea
A^{(0)}(p_1,p_2,p_3) & = & - \, 4 \, g_d \, 
{\rm Tr}(\lambda^a\lambda^b\lambda^c)~\Big(
\ve_1\cdot\ve_2\,p_2\cdot\ve_3 \nonumber \\
& & + \, \ve_2\cdot\ve_3\,p_3\cdot\ve_1
+ \ve_3\cdot\ve_1\,p_1\cdot\ve_2 +
O(\a') \Big)~~~,
\label{threetree}
\ena
and the four-gluon amplitude
\beqa
& & \! \! \! \! \! \! \! \! A_{4}^{(0)} (p_1 , p_2 , p_3 , p_4 ) = 
4 g_{d}^{2} \, 
Tr (\lambda^{a_1} \lambda^{a_2} \lambda^{a_3} \lambda^{a_4}) \, 
\frac{\Gamma (1 - \a' s) \Gamma (1 - \a' t)}{\Gamma (1 + \a ' u) \, s \, t} 
\label{fourtree} \\ 
& \times & \! \! \left[(\ve_1 \cdot \ve_2 ) (\ve_3  \cdot \ve_4 ) \, t \, u
+ (\ve_1  \cdot \ve_3 ) (\ve_2  \cdot \ve_4 ) \, t \, s +
(\ve_1  \cdot \ve_4 ) (\ve_2  \cdot \ve_3 ) \, s \, u +  \ldots \right]~,
\nonumber
\eeqa
where we have not written explicitly terms of the form $(\ve \cdot \ve)
(\ve \cdot p) (\ve \cdot p)$ and higher orders in $\a'$.
 
At one loop ($h=1$) we keep the gluons off the mass 
shell, and \eq{hmaster} gives, for $M~\geq~2$ transverse gluons,
\bea
& & A^{(1)}_P (p_1,\ldots,p_M) = N \, {\rm Tr}(\lambda^{a_1}
\cdots \lambda^{a_M}) \, \frac{g_d^M}{(4\pi)^{d/2}} \,
(2\a')^{(M-d)/2} (-1)^{M}  \nl
& \times & \int_{0}^{\infty} d \tau  {\rm e}^{2\tau} \, \tau^{-d/2}
\prod_{n=1}^\infty \left(1 - {\rm e}^{-2 n \tau} \right)^{2-d} \,
\int_0^\tau d \nu_M \int_0^{\nu_M} d \nu_{M-1} \dots 
\int_0^{\nu_3} d \nu_2 \nl
& \times & \left\{\prod_{i<j} \left[ 
\sqrt{\frac{z_i\,z_j}{V'_i(0)\,V'_j(0)}}
\exp \left(G(\nu_{ij}) \right)
\right]^{2\a' p_i\cdot p_j} \right.  \label{onemaster} \\
& \times & \left. \exp \left[ \sum_{i\not=j}
\left(\sqrt{2\a'} p_j\cdot\ve_i 
\, \partial_i  G(\nu_{ij})+ {1\over 2} 
\ve_i\cdot\ve_j \, \partial_i \partial_j
G(\nu_{ij}) \right) \right] \right\}_{\rm m.l.}~~~,    \nonumber
\ena
where $\nu_{ij} \equiv \nu_j - \nu_i$, $\partial_i \equiv 
\partial/\partial\nu_i$ and $\tau$ is related to the period 
${\tilde{\tau}}$ of the annulus by the relation
\beq
\tau = -{\rm i}\pi{\tilde \tau}~~~.
\label{tau}
\eeq
Instead of the Koba-Nielsen variables $z_i$, we have used the real 
variables
\beq
\nu_i = - \frac{1}{2} \log z_i~~~,
\label{nui}
\eeq
while the Green function $G(\nu_{ij} )$ is given by
\beq
G( \nu_{ji}) =
\log\left[- 2 \pi{\rm i}{{\theta_1\left(\frac{\rm i}{\pi}
(\nu_j-\nu_i)|\frac{\rm i}{\pi}\tau)\right)}\over{\theta'_1\left(0|
\frac{\rm i}{\pi}\tau\right)}}\right] - 
\frac{(\nu_j-\nu_i)^2}{\tau}~~~,
\label{onegreen}
\eeq
where $\theta_1$ is the first Jacobi $\theta$ function.

If we enforce the mass-shell condition $p_{i}^{2} =0$, any dependence 
on the local coordinates $V_i$'s drops out. However, in order to avoid 
infrared divergences, we will continue the gluon momenta off shell, in 
an appropriate way to be discussed later. Then, following Ref.~\cite{paper},
we will regard the freedom of choosing $V_i$ as a gauge freedom.
We make the simple choice
\beq
V_i'(0) = z_i~~~,
\label{Viprime}
\eeq
which will lead, in the field theory limit, to the background field 
Feynman gauge.
The conditions (\ref{Vi}) and (\ref{Viprime}) are easily satisfied by choosing
for example
\beq
V_i(z) = z_i \, z + z_i~~~.
\label{gaugech}
\eeq

\sect{The two-gluon amplitude}
\label{twopoint}

The one-loop two-gluon amplitude is given by
\beq
A^{(1)}(p_1,p_2) = N \, {\rm Tr}(\lambda^a\lambda^b) \,
\frac{g_d^2}{(4\pi)^{d/2}}(2\a')^{2-d/2}   
\ve_1 \cdot \ve_2  p_1 \cdot p_2 \, R( p_1 \cdot p_2)~,
\label{twoone}
\eeq
where 
\beq
R(s) = \int_0^\infty d\tau~{\rm e}^{2\tau}\,
\tau^{-d/2} \prod_{n=1}^\infty \left(1-{\rm e}^{-2n\tau} \right)^{2-d}
\int_0^\tau d\nu {\rm e}^{2\a' s \,G(\nu)} \, \left[ \partial_\nu G(\nu)
\right]^2~.
\label{Rint}
\eeq

Notice that if the two gluons are on mass shell,
the two-gluon amplitude becomes ill defined, as the kinematical
prefactor vanishes, while the integral diverges. In order to avoid this problem
we keep the two gluons off shell.

To take the field theory limit, we must remember 
that the modular parameter $\tau$ and the coordinate $\nu$ are related to 
proper-time Schwinger parameters for the Feynman diagrams contributing to the
two point function. In particular, $t \sim \a' \tau$ and $t_1 \sim \a' \nu$, 
where $t_1$ is the proper time associated with one of the two internal 
gluon propagators, while $t$ is the total proper time around the loop. 
In the field theory limit
these proper times have to remain finite, and thus the limit $\a' \to 0$
must correspond to the limit $\{ \tau, \nu \} \to \infty$ in the integrand.
The field theory limit is then determined by
the asymptotic behavior of the Green function for large $\tau$, namely
\beq
G(\nu, \tau) = - \frac{\nu^2}{\tau} + \log\left(2\sinh(\nu)\right)
- 4 \, {\rm e}^{-2\tau} \, \sinh^2(\nu) + 0 ( {\rm e}^{-4 \tau})~~~, 
\label{lartauG}
\eeq
where $\nu$ must also be taken to be large, so that $\hat\nu$ remains finite;
in this region, we may use
\beq
G(\nu, \tau) \sim  (\hat\nu - \hat\nu^2) \tau -
{\rm e}^{- 2 \hat\nu \tau} - {\rm e}^{- 2 \tau (1 - \hat\nu)}
+ 2 {\rm e}^{- 2 \tau}~~~,
\label{lartaunuG}
\eeq
so that
\beq
\frac{\partial G}{\partial \nu} \sim 1 - 2 \hat\nu +
2 {\rm e}^{- 2 \hat\nu \tau} - 2 {\rm e}^{- 2 \tau (1 - \hat\nu)}~~~.
\label{lartaudG}
\eeq

We now substitute these results into \eq{twoone}, and keep only terms that 
remain finite when $k=e^{-2\tau}\to 0$. Divergent terms must be discarded by 
hand, since they correspond to the propagation of the tachyon in the loop. 
The next-to-leading term corresponds to gluon exchange, and while it is also
divergent in the field theory limit, the corresponding divergence
is regularized by dimensional regularization. Finally,
higher order terms $ {\rm e}^{-2n \tau} $ with $n>0$ are vanishing in 
the field theory limit.

Notice that by taking the large $\tau$ and $\nu$ limit we have
discarded two singular regions of integration that potentially contribute in
the field theory limit, namely $\nu \to 0$ and $\nu \to \tau$. 
In these regions (often referred to as ``pinching'' regions)
the Green function has a logarithmic singularity corresponding to the
insertion of the two external states very close to each other, and this
singularity in general gives 
non-vanishing contributions in the field theory limit. 
However, in the case of the two gluon amplitude, 
these regions correspond to Feynman diagrams with a loop
consisting of a single propagator, {\it i. e.} a ``tadpole''. Massless
tadpoles are defined to vanish in dimensional regularization, and thus we are
justified in discarding these contributions as well.

Replacing the variable $\nu$ with ${\hat \nu}\equiv\nu/\tau$, 
\eq{Rint} becomes
\beq
R(s) = \int_0^\infty d\tau \int_0^1 d{\hat \nu} ~\tau^{1-d/2}\,
{\rm e}^{2\a'\,s\,({\hat\nu}-{\hat\nu}^2)\tau}
\left[(1-2{\hat\nu})^2(d-2)-8\right]~~~,
\label{limRint}
\eeq
so that the integral is now elementary, and yields
\beq
R(s) = - \Gamma\left(2-\frac{d}{2}\right)\, (- 2 \a' s)^{d/2-2}~
\frac{6-7d}{1-d}\,B\left(\frac{d}{2}-1,\frac{d}{2}-1\right)~~~,
\label{Rfin}
\eeq
where $B$ is the Euler beta function.

If we substitute \eq{Rfin} into \eq{twoone}, we see that the
$\a'$ dependence cancels, as it must. The ultraviolet finite
string amplitude, \eq{twoone}, has been replaced by a field
theory amplitude which diverges in four space-time dimensions, 
because of the pole in the $\Gamma$ function in \eq{Rfin}.
Defining as usual a dimensionless coupling 
constant $g_d = g\,\mu^\e$, with $\mu$ an arbitrary mass scale,
and having set $d=4-2\e$, we find 
\beq
A^{(1)}(p_1,p_2) = -N \frac{g^2}{(4\pi)^2} \,
\left(\frac{4\pi\,\mu^2}{-p_1\cdot p_2}\right)^\e \,
\Gamma(\e)\,\frac{11-7\e}{3-2\e}\,
B(1-\e,1-\e) A^{(0)} (p_1 , p_2 )
\label{twofin}
\eeq
\eq{twofin}) is exactly equal to the gluon vacuum
polarization of the $SU(N)$ gauge field theory that one computes
with the background field method, in Feynman gauge, with dimensional 
regularization, provided we use for the tree-level two-gluon 
amplitude the expression
\beq
A^{(0)} (p_1 , p_2 ) =  \delta^{ab} \left[  \ve_1\cdot
\ve_2\,p_1\cdot p_2 - \ve_1 \cdot p_2 \,\, \ve_2 \cdot
p_2 \right]
\label{treetwogluon}
\eeq

Comparing \eq{twofin} with the equation for $\Gamma_2$ in Eq. (\ref{renprop1})
we can extract the minimal subtraction wave function renormalization 
constant
\beq
Z_A= 1 + N\,\frac{g^2}{(4\pi)^2}\,\frac{11}{3}\,\frac{1}{\e}~~~.
\label{za}
\eeq
While this result is what we expected, it relies on our prescription
to continue the string amplitude off shell, and on our choice of the 
projective transformations $V_i$. To make sure that our prescription
is consistent we need to compute the three and four point
renormalizations as well, and verify that gauge invariance is preserved.

\sect{An alternative computation of proper vertices}
\label{effact}

In the previous section we have computed the 1PI two-gluon
amplitude and we have extracted the wave function renormalization constant. 
In this section we present an alternative method, introduced by Metsaev and 
Tseytlin~\cite{mets}.
This method isolates the 1PI part of the amplitude, and is
thus particularly suited to the evaluation of renormalization constants. It 
is based on the following  alternative expression for the bosonic Green 
function~\cite{fratse1} 
\beq
G(\nu_i, \nu_j) = - \sum_{n=1}^{\infty}~\frac{1 + q^{2n}}{n (1 - q^{2n})} 
\cos 2 \pi n \left( \frac{\nu_j - \nu_i}{\tau} \right) + ~\dots~~~,
\label{newmtgreen}
\eeq
where $q = e^{- \pi^2/\tau}$ and
the dots stand for terms independent of $\nu_i$ and $\nu_j$, that 
will not be important in our discussion.
 
An important advantage of this approach is that, at least at one loop, it 
allows to integrate exactly over the punctures before the field theory limit
is taken. The result does not present pinching singularities, that are 
regularized directly in the Green function. As a consequence, for the 
two gluon amplitude, we will get the same expression that we derived
in \secn{twopoint}, while for the three and four gluon amplitudes 
we will get only the contributions that do not include pinchings and are
therefore one-particle irreducible. 

As a first step, we rewrite the one-loop $M$-gluon planar amplitude as
\beqa
A^{(1)}_P (p_1,\ldots,p_M) & = & N \, {\rm Tr}(\lambda^{a_1}
\cdots \lambda^{a_M}) \,
\frac{g_d^M}{(4\pi)^{d/2}} \, (2 \a')^{2 - d/2} \label{newonemast} \\
& \times & \! \! (- 1)^M \int_{0}^{\infty} d \tau \, {\rm e}^{2 \tau}\, 
\tau^{-d/2}
\prod_{n=1}^{\infty} \left( 1 - {\rm e}^{-2n \tau} \right)^{2-d} 
I^{(1)}_M(\tau)~,
\nonumber
\eeqa
where $I^{(1)}_M(\tau)$ is the integral over the punctures $\nu_i$, and can
be read off from \eq{onemaster}.

For $M=2$, after a partial integration with vanishing surface term, we get
\beq
I^{(1)}_2(\tau) = p_1 \cdot p_2 \, \ve_1 \cdot \ve_2 
\int_{0}^{\tau} d \nu
\left( \partial_\nu G(\nu) \right)^2 \left(e^{G(\nu)}\right)^{2 
\a' p_1 \cdot p_2}~~~.
\label{2int}
\eeq
Since we are only interested in divergent renormalizations, and since the
overall power of $\a'$ is already appropriate to the field theory limit, 
as it vanishes when $d \to 4$, we can now neglect the exponential, which would 
contribute $1 + O(\a')$. 
Using the expression in \eq{newmtgreen} for the Green 
function, we can easily perform exactly the integral over the puncture, and
we get
\beq
I^{(1)}_2(\tau) = \frac{2 \pi^2}{\tau} \, p_1 \cdot p_2 \ve_1 
\cdot \ve_2 \, \sum_{n=1}^{\infty} \left(
\frac{1 + q^{2n}}{1 - q^{2n}} \right)^2~~~,
\label{2res}
\eeq
so that we can write
\beqa
A^{(1)}(p_1, p_2) & = & \frac{N}{2} {\rm Tr} \left( \lambda^{a_1}
\lambda^{a_2} \right) 
\frac{g_d^2}{(4 \pi)^{d/2}} (2 \a')^{2 - d/2}  
p_1 \cdot p_2 \ve_1 \cdot \ve_2 \, Z(d)  \nl
& = & \frac{N}{4} \, \frac{g_d^2}{(4 \pi)^{d/2}} \, ( 2 \alpha ')^{2 -d/2} 
\, Z(d) \, A^{(0)}(p_1, p_2)~~~.
\label{2onetree}
\eeqa
Here
\beq
Z(d) \equiv (2 \pi)^2 \int_0^\infty d \tau \, {\rm e}^{2 \tau} \,
\tau^{-1-d/2} \, \prod_{n=1}^{\infty} \left( 1 - {\rm e}^{-2n \tau} 
\right)^{2-d}
\sum_{m=1}^{\infty}
\left(\frac{1 + q^{2m}}{1- q^{2m}}\right)^2
\label{Zint}
\eeq
is the string integral that generates the renormalization constants as 
$\a' \to 0$.

With three gluons we get
\beqa
I^{(1)}_3(\tau) & = & \int_{0}^{\tau} d \nu_3 \int_{0}^{\nu_3} 
d \nu_2 \left\{ \ve_1 \cdot \ve_2 \, \partial_{1} \partial_2 
G(\nu_{21})  \right.  \nl
& \times & \left. \left[ p_1 \cdot \ve_3 \, \partial_3 G (\nu_{31}) +
p_2 \cdot \ve_3 \, \partial_3 G(\nu_{32}) \right] 
+ \dots \right\}~~~, 
\label{3int}
\eeqa
where terms needed for cyclic symmetry and terms of order $\a'$ are not 
written explicitly, and we discarded the exponentials of the Green
functions, that are not contributing since the external gluons are on shell.
 
The integrals over $\nu_2$ and $\nu_3$ can be done by using the expression 
in \eq{newmtgreen} for the Green function. The result is 
\beqa
I^{(1)}_3(\tau) & = & \frac{(2 \pi)^2}{\tau}
\left[\ve_1 \cdot \ve_2 p_2 \cdot \ve_3 +
\ve_2 \cdot \ve_3 p_3 \cdot \ve_1 +
\ve_1 \cdot \ve_3 p_1 \cdot \ve_2 \right]  \nl
& \times & \sum_{n=1}^{\infty} 
\left( \frac{1+ q^{2n}}{1 - q^{2n}} \right)^2 \, + \, 0(\a')~~~,
\label{3res}
\eeqa
so that the three gluon amplitude is given by
\beq
A^{(1)}(p_1, p_2, p_3) = \frac{N}{4} 
\frac{g_d^2}{(4 \pi)^{d/2}} (2 \a')^{2 - d/2} Z(d) A^{(0)}(p_1, p_2, p_3)
\, + \, O(\a')~~~.
\label{3onetree}
\eeq
 
Finally, the same calculation can be done for the four-gluon amplitude,
where we can concentrate on the terms whose kinematical prefactor has
no powers of the external momenta (and thus is of the form 
$\ve_i \cdot \ve_j \, \ve_h \cdot \ve_k$). Other terms are suppressed
by powers of $\a'$. Then we need to consider the expression
\beqa
I^{(1)}_4(\tau) & = & \int_0^\tau d \nu_4 \int_0^{\nu_4} d \nu_3 
\int_0^{\nu_3} d \nu_2 \Big[ \ve_1 \cdot \ve_2 \, 
\ve_3 \cdot \ve_4 \, \partial_1 \partial_2 G(\nu_{21}) \, 
\partial_3 \partial_4 G(\nu_{43})  \nl
& + & \ve_1 \cdot \ve_3  \, 
\ve_2 \cdot \ve_4 \, \partial_1 \partial_3 G(\nu_{31}) \,
\partial_2 \partial_4 G(\nu_{42})  \label{4int}  \\
& + & \ve_1 \cdot \ve_4  \, 
\ve_3 \cdot \ve_2 \, \partial_1 \partial_4 G(\nu_{41}) \,
\partial_3 \partial_2 G( \nu_{32}) \, \Big]~~~.
\nonumber
\eeqa
Using again \eq{newmtgreen}, we can perform the integrals over the punctures,
and we get
\beqa
& & I^{(1)}_4(\tau) = \frac{(2 \pi)^2}{\tau} \sum_{n=1}^{\infty} 
\left(\frac{1 + q^{2n}}{1 - q^{2n}} \right)^2 \label{4res} \\
& \times & \left[\ve_1 \cdot \ve_3  \,
\ve_2 \cdot \ve_4 - \frac{1}{2}
\ve_1 \cdot \ve_2  \, 
\ve_3 \cdot \ve_4  - \frac{1}{2}
\ve_2 \cdot \ve_3  \, 
\ve_1 \cdot \ve_4 \right]~~~.
\nonumber
\eeqa
The amplitude becomes then
\beq
A^{(1)}(p_1, p_2, p_3, p_4) = \frac{N}{4} 
\frac{g_d^2}{(4 \pi)^{d/2}} (2 \alpha ')^{2 - d/2} Z(d) 
A^{(0)}(p_1, p_2, p_3, p_4) \, + \, O(\a')~~~,
\label{4onetree}
\eeq
where the 1PI part of the four-gluon amplitude at
tree level is given by
\beqa
A^{(0)}(p_1, p_2, p_3, p_4) & = & 4 \, g_{d}^{2} \,  
Tr ( \lambda^{a_1} \lambda^{a_2} \lambda^{a_3} \lambda^{a_4} ) 
\label{fourone} \\
& \times &
\left[ \e_1 \cdot \e_3 \, 
\e_2 \cdot \e_4  - \frac{1}{2} \e_1 \cdot \e_2 
\e_3 \cdot \e_4 - \frac{1}{2} \e_2 \cdot \e_3 \,
\e_1 \cdot \e_4 \right]~~~.
\nonumber
\eeqa

Defining the factor  
\beq
K(d) = \frac{N}{4} \,
\frac{g_d^2}{(4 \pi)^{d/2}} \, (2 \a')^{2 -d/2} \,  Z(d)~~~,
\label{loopefflag}
\eeq
we can now perform the limit $\a' \to 0$, keeping the ultraviolet 
cutoff $\e \equiv 2 - d/2$ small but positive, and eliminating by hand 
the tachyon contribution. The calculation of the integral $Z(d)$ in this
limit is described in detail in Ref.~\cite{paper}. The result is 
\beq
K(4 - 2 \e) \rightarrow
- \frac{11}{3} \, N \, \frac{g^2}{(4\pi)^2} \, \frac{1}{\e} \,
+ \, O(\e^0)~~~.
\label{divefflag}
\eeq

If we finally compare Eqs. (\ref{renprop1}) with Eqs. (\ref{2onetree}), 
(\ref{3onetree}) and (\ref{4onetree}) we can determine
the renormalization constants. They are given by
\beq
Z_A = Z_3 = Z_4 = 1   + \frac{11}{3} \, N \, \frac{g^2}{(4 \pi)^2} \, 
\frac{1}{\e}~~~,
\label{wardagain}
\eeq
in agreement with the result of the previous section for $Z_A$, and as 
dictated by the background field method Ward identities.

\sect{The three-gluon amplitude}
\label{threepoint}

The methods described in the previous two sections are both adequate to
compute one-particle irreducible contributions to the Green functions. 
Reducible diagrams, on the other hand, correspond to regions in moduli space
where the gluons are inserted on the string world sheet very close to each 
other (pinching regions). These regions were excluded by hand in \secn{effact},
since the corresponding logarithmic singularity in the world-sheet Green 
function was regularized by a $\zeta$-function regularization~\cite{paper}. 
If we wish to
include them along the lines of \secn{twopoint}, we have to perform the
field theory limit in a slightly different way. To see this, let us consider 
the simplest case in which these contributions arise, namely the three-gluon
amplitude. 

The one-loop correction to \eq{threetree} can be written as 
\bea
A^{(1)}(p_1,p_2,p_3) & = & - N \, {\rm Tr}(\lambda^a\lambda^b\lambda^c) \,
\frac{g_d^3}{(4\pi)^{d/2}} \, (2\a')^{2-d/2} \nl
& \times & \!\!\! \int_0^\infty {\cal D}\tau \int_0^\tau \!d\nu_3 
\int_0^{\nu_3}\!d\nu_2 ~f_3(\nu_2,\nu_3,\tau)~~~,
\label{threeone}
\ena
where
\bea
f_3(\nu_2,\nu_3,\tau) & \equiv &
{\rm e}^{2\a'p_1\cdot p_2\,G(\nu_{2})}~
{\rm e}^{2\a'p_2\cdot p_3\,G(\nu_{32})}~
{\rm e}^{2\a'p_3\cdot p_1\,G(\nu_3)} \nl
& \times & \Bigg\{ \left[
- \, \ve_1\cdot\ve_2 \, \partial_2^2 G(\nu_2) \left(
p_1\cdot\ve_3 \, \partial_3 G(\nu_3) + 
p_2\cdot\ve_3 \, \partial_3 G(\nu_{32}) \right) \right.  \nl
& & + \left. \, \ve_2\cdot\ve_3 \, \partial_3^2 G(\nu_{32}) 
\left(p_2\cdot\ve_1 \, \partial_2 G(\nu_2) +
p_3\cdot\ve_1 \, \partial_3 G(\nu_3) \right) \right.  \nl
& & + \left. \, \ve_1\cdot\ve_3 \, \partial_3^2 G(\nu_3) \left(
p_3\cdot\ve_2 \, \partial_3G(\nu_{32}) -
p_1\cdot\ve_2 \, \partial_2G(\nu_2) \right) \right]  \nl
& & + \, \, O(\a') \Bigg\}~~~,
\label{threeint}
\ena
and
\beq
{\cal D}\tau \equiv d\tau~{\rm e}^{2\tau}\,\tau^{-d/2}
\prod_{n=1}^\infty \left(1-{\rm e}^{-2n\tau} \right)^{2-d}~~~.
\label{taumeas}
\eeq

One-particle irreducible contributions can be calculated along the lines of
\secn{twopoint}, expanding the bosonic world-sheet Green
function for large values of $\tau$ as in \eq{lartauG}. The calculation is
described in some detail in Ref.~\cite{paper}, and gives
\beq 
A^{(1)}(p_1,p_2,p_3)\Big|_{\rm 1PI} =
\left(-{11\over 3}\right) \,N \, \left({g\over 4
\pi}\right)^2 {1\over\e} \, A^{(0)}(p_1,p_2,p_3)
+ O(\e^0)~~~,
\label{1pi3}
\eeq
which agrees with the results of \secn{effact}.

Next, we turn to the analysis of the pinching regions. There are
clearly three such regions, corresponding
to $\nu_2 \to 0$, $\nu_2 \to \nu_3$ and $\nu_3 \to \tau$, as dictated
by cyclic symmetry and periodicity on the annulus.

Let us consider, for example, the first region, $\nu_2 \to 0$. 
Since this pinching contribution is localized in a neighbourhood
of $0$, we can replace the integral $\int_0^{\nu_3} d \nu_2$ with
an integral $\int_0^\eta d \nu_2$, where $\eta$ is an arbitrary small
number. Further, we can use for the bosonic Green function the
approximation 
\beq
G(\nu) \sim \log(2 \, \nu)~~~.
\label{pinchG}
\eeq
After this is done, in $f_3(\nu_2,\nu_3,\tau)$ we can expand
$G(\nu_{32})$ in powers of $\nu_2$, which turns the amplitude 
$A^{(1)}(p_1,p_2,p_3)$ into an infinite series. 
The $n$-th term of this series is proportional to an integral of the form
\beq
C_n \equiv \int_0^\eta d\nu_2~\nu_2^{n-2+2\a'p_1\cdot p_2}~~~,
\label{pinchser}
\eeq
with $n \geq 0$. After a suitable analytic continuation in the momenta 
to insure convergence, we get
\beq
C_n = \frac{\eta^{n-1+2\a'p_1\cdot p_2}}{n-1+2\a'p_1\cdot p_2}~~~.
\label{pinchpol}
\eeq
We see that, when the pinching $\nu_2\to 0$ is performed, the
amplitude becomes an infinite sum over
all possible string states that are exchanged in the $(12)$-channel, 
$n=0$ corresponding to the tachyon, $n=1$ to the gluon and so on.
In the case of the three-gluon amplitude, one can verify that
the exchange of a tachyon does not give any contribution: in fact 
the coefficient of the quadratic divergence ${1\over\nu_2^2}$ is
zero because of the trasversality of the externals states. 
The gluon contribution, on the other hand, survives in the field theory
limit, and contributes to the ultraviolet divergence, as expected: the single
pole in $\nu_2$ in fact generates, through the change of variable to
$\hat\nu_2$, the negative power of $\tau$ needed for the integral to diverge. 
All other terms in the series, corresponding to $n \geq 2$, and to states 
whose mass becomes infinite as $\a' \to 0$, vanish in the field theory 
limit. Notice also that for $n = 1$ the dependence on the cutoff $\eta$
in \eq{pinchpol} disappears as $\a' \to 0$.

Keeping this in mind, and collecting all relevant factors, we find that 
the pinching contribution to the three gluon amplitude that we are considering 
is
\bea
A^{(1)}(p_1,p_2,p_3)\Big|_{\nu_2\to 0} & = &
- \, N \, {\rm Tr}(\lambda^a\lambda^b\lambda^c) \,
\frac{g_d^3}{(4\pi)^{d/2}} \, (2\a')^{2-d/2} \label{pinch} \\
& \times & \frac{(p_1+p_2)\cdot p_3}{p_1\cdot p_2}~
R\left[(p_1+p_2)\cdot p_3\right] \,
\ve_1\cdot\ve_2 \, p_2\cdot\ve_3~~~, \nonumber 
\ena
where $R$ is the integral defined in \eq{Rint}. Notice that
\eq{pinch} contains a ratio of momentum invariants which are vanishing
on shell. The appearance of such ratios in string amplitudes, in the
corners of moduli space corresponding to loops isolated on external legs, is
a well-known fact, which for example motivated the work of 
Ref.~\cite{berkosrol}. As we already remarked in Ref.~\cite{letter}, 
this ``$0/0$'' ambiguity is similar to the one that appears in 
the unrenormalized connected
Green functions of a massless field theory, if the external legs
are kept on the mass-shell and divergences are regularized with dimensional
regularization. 
Our prescription to deal with this ambiguity is to continue off shell the  
momentum of the gluon attached to the loop, according to
\beq 
p_3^2 = (p_1 + p_2)^2 = m^2~~~.
\label{off3}
\eeq
The other gluon momenta, $p_1$ and $p_2$, on the other hand, are kept on shell.
Here we rely on the assumption,
subtantiated by the results obtained so far, that string amplitudes lead
to field theory amplitudes calculated with the background field method.
As was shown in Ref.~\cite{abbgrisch}, $S$-matrix elements
are obtained in this method by first calculating one-particle irreducible 
vertices to the desired order, and then gluing them together with 
propagators that are defined by fixing the gauge for the 
background field.
This leads us to interpret \eq{pinch}
as a one-loop, one-particle irreducible two point function, whose
momentum must be continued off shell according to \eq{off3}, glued to a 
tree-level three point vertex, for which no off-shell continuation is
necessary. We thus keep $p_1^2 = p_2^2 = 0$, which, using momentum 
conservation, implies
\beq
p_1\cdot p_2 = \frac{m^2}{2}~~~. 
\label{off12}
\eeq
Then, comparing \eq{pinch} with \eq{threetree}, and including a factor
of three to account for the three pinching regions, we can write
\beq
A^{(1)}(p_1,p_2,p_3)\Big|_{pinch.} = - \, \frac{3}{2} \, N 
\, \frac{g_d^2}{(4\pi)^{d/2}} \, (2\a')^{2-d/2} \, R(- m^2) \,
A^{(0)}(p_1,p_2,p_3)~~~.
\label{pincha0}
\eeq

Extracting the ultraviolet divergence of \eq{pincha0}, and adding 
\eq{1pi3}, we find the total divergence of the unrenormalized, connected, 
three gluon Green function,
\beq 
A^{(1)}(p_1,p_2,p_3)\Big|_{\rm div} = 2 \left({11\over 3}\right)
\, N \, \left({g\over 4\pi}\right)^2 {1\over\e} \,
A^{(0)}(p_1,p_2,p_3)~~~,
\label{totdiv3}
\eeq
which leads again to the background field Ward identity ( see Eq. 
(\ref{renprop2}))
\beq 
Z_3 = Z_A = 1 + N \left({g\over 4\pi}\right)^2 {11\over 3}
({1\over\e})~~~.
\label{z3=za}
\eeq

The same analysis can be carried out for the four-point amplitude, as described
in Ref.~\cite{paper}, and no surprises arise.

\sect{Concluding remarks}
\label{concl}

We have shown that it is possible to calculate renormalization constants in
Yang-Mills theories using the simplest of string theories, the open bosonic
string. To do so it is necessary to continue off shell some of the external
momenta, but this can be done consistently in the field theory limit, and 
the results concide with the ones obtained using the background field method
and dimensional regularization. Since bosonic string amplitudes are
well understood at all orders in perturbation theory, this technique
may be useful beyond one loop.

\end{document}